\documentclass[aps,pra,twocolumn,amsmath,amssymb]{revtex4}
\usepackage{graphicx,epsfig,epsf}% Include figure files
\usepackage{dcolumn}% Align table columns on decimal point
\usepackage{bm}% bold math

\begin{document}
\sloppy

\title{Role of interference in quantum state transfer through spin chains}

\author{A. O. Lyakhov$^{(1)}$, D. Braun$^{(2)}$, and C.
  Bruder$^{(1)}$}
\affiliation{$^{(1)}$Department of Physics and Astronomy, University of Basel,
  Klingelbergstr. 82,
4056 Basel, Switzerland\\
$^{(2)}$Laboratoire de Physique Th\'eorique, IRSAMC, UMR 5152 du CNRS,
Universit\'e Paul Sabatier, 118, route de Narbonne, 31062 Toulouse, FRANCE}

\begin{abstract}
We examine the role that interference plays in quantum state transfer
through several types of finite spin chains, including chains with
isotropic Heisenberg interaction between nearest neighbors, chains
with reduced coupling constants to the spins at the end of the chain, and
chains with anisotropic coupling constants. We evaluate quantitatively both the
interference corresponding to the propagation of the entire chain, and
the interference in the effective propagation of the first and last
spins only, treating the rest of the chain as black box. We show that
perfect quantum state transfer is possible without quantum
interference, and provide evidence that the spin chains examined
realize interference-free quantum state transfer to a good
approximation.
\end{abstract}
%\pacs{03.67.-a, 03.67.Lx, 03.67.Mn }
\maketitle
%\begin{twocolumn}

\section{Introduction}
Recently, the idea to use quantum spin chains for short-distance quantum
communication was put forward by Bose \cite{Bose03}. He showed that an array of
spins (or spin-like two level systems) with isotropic Heisenberg interaction is
suitable for quantum state transfer.  In particular, spin chains can be used as
transmission lines for quantum states without the need to have controllable
coupling constants between the qubits or complicated gating schemes to achieve
high transfer fidelity.  Alice prepares an input state on the first spin of the
chain at time $t=0$ (all other spins are in the state down/zero), and after a
certain time $t_{1}$ Bob recovers an output state on the last spin at the other
end, while Alice normally looses her state, in fulfillment of the no-cloning
theorem \cite{Wootters82}.  Bose showed that the average fidelity of the
quantum state transfer exceeds the maximum value which can be achieved
classically for spin chains of length $N$ up to $N\sim 80$. The method was
later extended using engineered coupling constants
\cite{Christandl04,Yung04,Lyakhov06}, multi-spin encoding \cite{Osborne04}, as
well as ``dual-rail encoding'' in two parallel quantum channels
\cite{Burgarth04,Burgarth05}. In \cite{Wojcik05} it was shown that the
transmission even through very long chains can be improved to almost perfect
fidelity if the coupling of the first and last spins to the chain is reduced.

As with any task in quantum information processing which offers an
advantage over classical information processing, the question arises
what in the quantum world allows for that advantage. It is generally
acknowledged that quantum entanglement and interference are two
ingredients which distinguish quantum information processing from its
classical counterpart \cite{Bennett00}. Quantum entanglement has been
studied in great detail over the last fifteen years
\cite{Lewenstein00}, but the precise role of interference in various
quantum information treatment tasks remains to be elucidated
\cite{Beaudry05}.

Contrary to entanglement, interference is a property not of a quantum
state but of the propagator of a state. This is due to the fact that
the coherence of the propagation is important for
interference. Indeed, the final probability distribution resulting
from a given quantum algorithm can always be generated through
stochastic simulation on a classical computer as well: for a known
quantum circuit and initial state one can in principal calculate the
final state, and therefore the probability distribution. It is then
simple to create a stochastic process which gives each possible
outcome with the correct probability.  In such a classical simulation
clearly no interference takes place. Thus, what counts for
interference is not a state itself but the way it was created.

A quantitative measure of interference in any quantum mechanical
process in a finite-dimensional Hilbert space was recently introduced
in \cite{Braun06}, and the statistics of quantum interference in
random quantum algorithms was studied in \cite{Arnaud06}. Here we
propose to study the role of interference in quantum state transfer
through spin chains.  After defining the notion of interference,
reduced interference, and fidelity in Section
\ref{section_def_interference}, we will follow two complementary
approaches: in Section \ref{section_reduced_interference} we will
consider the spin chain as a black box which propagates the initial
state of the first and last spins combined to a final state of these
two spins. We will calculate the reduced interference that describes
this propagation for different spin chains and show that perfect state
transfer is possible without quantum interference. In Section
\ref{section_full interference} we will then consider the unitary
evolution of the entire chain and analyze this unexpected result.

\section{Interference and reduced interference}
\label{section_def_interference}
The interference $I(t)$ for a general quantum process described by a
propagator $\mathcal{P}(t)=P_{ij,kl}(t)$ which propagates
an initial state $\rho$ with matrix elements $\rho_{ij}$ in a fixed
orthonormal basis of dimension $D$ to a final state $\rho'$,
\begin{equation}\label{rho'}
\rho^\prime_{ij}=\sum_{k,l=1}^D P_{ij,kl}\rho_{kl}
\end{equation}
is defined as \cite{Braun06}
\begin{equation}\label{IP}
I(t)=\sum_{i,k,l}|P_{ii,kl}(t)|^2-\sum_{i,k}|P_{ii,kk}(t)|^2\,.
\end{equation}

If $\mathcal{P}$ describes the propagation of the reduced density
matrix of the first and last spins alone (which will be mixed in general,
as it results from tracing out the intermediate spins of the chain),
Eq.~(\ref{IP}) defines the ``reduced interference'' $I_r(t)$. We will
evaluate $I_r(t)$ analytically for spin chains which conserve the
number of excitations in the chain, and show that $I_r(t)$ is
intimately linked to the average fidelity $F(t)$ introduced in
\cite{Bose03},
\begin{equation}\label{ft}
F(t)=\frac{1}{4\pi}\int \langle \psi_{in}|\rho_{out}(t)|
\psi_{in}\rangle d\Omega\,,
\end{equation}
where $|\psi_{in}\rangle$ is the pure state to be transmitted
prepared on the first
spin, $\rho_{out}$ is the output state on the last spin
(i.e.~$\rho_{out}=\mathrm{Tr}_1\rho'$, with the trace over the first (input)
spin), and the integral is over all initial states of the input spin on the
Bloch sphere parameterized by the spatial angle $\Omega$. $F(t)$ is a function
of time and we are interested in the maximum fidelity in some reasonable time
interval, that is less than the decoherence time of the qubits, and scales like
the inverse coupling \cite{Bose03}. We will also provide numerical results for
$I_r(t)$ for chains in which the number of excitations is not conserved.

The interference measure for the unitary propagation of the entire chain,
$|\psi'\rangle=U|\psi\rangle$ reduces to \cite{Braun06}
\begin{equation}
I_U(t)=D-\sum_{i,k=1}^D|U_{i,k}(t)|^4\,.
\label{IU}
\end{equation}
For this coherent propagation, the interference $I_U(t)$ measures the degree of
equipartition of the output that result from any basis state of a system at
$t=0$. Here, an equipartitioned state means a state that is a superposition of
all the basis states with amplitudes of modulus $1/\sqrt{D}$. For better
comparison of the results we will plot the normalized interference
$I=I_U/(D-1)$ so that the maximal possible value of the interference is one and
does not depend on the number of qubits in the chain.

\section{Reduced interference for excitation-conserving spin chains}
\label{section_reduced_interference}
We start by evaluating the reduced interference in excitation-preserving
chains, i.e., spin chains for which the total Hamiltonian $H$
commutes with the total spin component $S^z=\sum_{i=1}^N\sigma_i^z$. The
particular example of the chain with isotropic Heisenberg interaction proposed
by Bose \cite{Bose03} falls into this class (see Section \ref{sec.Bosechain}
below). We start with at most one excitation in the chain and limit ourselves
to pure initial states. Therefore one can specify a state of the entire chain
$|j\rangle$ ($j=1,\ldots,N$) by the position at which the excitation is
localized. In principle there are four computational basis states for the two
spins, but the state where both the first and the last spins are excited will
never appear. We therefore restrict our attention to the three-dimensional
Hilbert space spanned by the states $|1\rangle$, $|N\rangle$, and $|0\rangle_r$
(the state where both the first and the last spins are not excited). We will
also make use of the state $|0\rangle_m$ of the intermediate part of the chain,
where all intermediate spins are not excited.

We start from an initial state of the chain which factorizes between the two
selected spins ($1$ and $N$) and the rest of the chain, which is
assumed to be in state
$|0\rangle_m$,
\begin{equation}
|\Psi_{in}\rangle=(a_0|0\rangle_r+a_1|1\rangle+a_N|N\rangle)|0\rangle_m\,.
\label{psin}
\end{equation}
The initial reduced density matrix of the first and last spins,
\begin{equation}
\rho=\left(
\begin{array}{ccc}
|a_0|^2 & a_0a_1^* & a_0a_N^* \\
a_1a_0^* & |a_1|^2 & a_1a_N^* \\
a_Na_0^* & a_Na_1^* & |a_N|^2
\end{array}\right)\,,
\label{rho1}
\end{equation}
then still represents a pure state. For any Hamiltonian that conserves the
number of excitations we can write the state at time $t$ as
\begin{widetext}
\begin{equation}
|\Psi_{out}(t)\rangle=a_0|0\rangle_r|0\rangle_m+a_1\sum_{j=1}^N\langle
j|e^{-iHt}|1\rangle|j\rangle+a_N\sum_{j=1}^N\langle
j|e^{-iHt}|N\rangle|j\rangle\,,\label{psit}
\end{equation}
or
\begin{eqnarray}
|\Psi_{out}(t)\rangle=a_0|0\rangle_r|0\rangle_m+a_1f_{11}|1\rangle|0\rangle_m+
a_1\sum_{j=2}^{N-1}\langle j|e^{-iHt}|1\rangle|0\rangle_r|j\rangle+
a_1f_{N1}|N\rangle|0\rangle_m+ \nonumber\\
a_Nf_{1N}|1\rangle|0\rangle_m+ a_N\sum_{j=2}^{N-1}\langle
j|e^{-iHt}|N\rangle|0\rangle_r|j\rangle+ a_Nf_{NN}|N\rangle|0\rangle_m\,,
\end{eqnarray}
where
\begin{equation}
f_{ij}(t)=\langle i|e^{-iHt}|j\rangle\,.
\end{equation}

After tracing out the intermediate spins we obtain the final density matrix
of the first and last spins,
\begin{equation}
%\rho_{1N}=\left(
\rho'=\left(
\begin{array}{ccc}
|a_0|^2+ S_m & a_0(a_1^*f_{N1}^*+a_N^*f_{NN}^*) & a_0(a_1^*f_{11}^*+a_N^*f_{1N}^*) \\
a_0^*(a_1f_{N1}+a_Nf_{NN}) & |a_1f_{N1}+a_Nf_{NN}|^2 & (a_1f_{N1}+a_Nf_{NN})(a_1^*f_{11}^*+a_N^*f_{1N}^*) \\
a_0^*(a_1f_{11}+a_Nf_{1N}) & (a_1f_{11}+a_Nf_{1N})(a_1^*f_{N1}^*+a_N^*f_{NN}^*)
& |a_1f_{NN}+a_Nf_{1N}|^2
\end{array}\right)
\label{rho2}
\end{equation}
where
\begin{equation}
S_m=\sum_{j=2}^{N-1} |a_1\langle j|e^{-iHt}|1\rangle+a_N\langle
j|e^{-iHt}|N\rangle|^2\,.
\end{equation}

Comparing Eqs.~(\ref{rho1}), (\ref{rho2}), and (\ref{rho'}) we read off the
propagator
\begin{equation}
\mathcal{P}=\left(
\begin{array}{ccccccccc}
1 & 0 & 0 & 0 & \sum_{j}|f_{j1}|^2 & \sum_{j}f_{j1}f_{jN}^* & 0 &\sum_{j}f_{jN}f_{j1}^* & \sum_{j}|f_{jN}|^2\\
0 & f_{N1}^* & f_{NN}^* & 0 & 0 & 0 & 0 & 0 & 0 \\
0 & f_{11}^* & f_{1N}^* & 0 & 0 & 0 & 0 & 0 & 0 \\
0 & 0 & 0 & f_{N1} & 0 & 0 & f_{NN} & 0 & 0 \\
0 & 0 & 0 & 0 & |f_{N1}|^2& f_{N1}f_{NN}^*& 0 & f_{NN}f_{N1}^*& |f_{NN}|^2\\
0 & 0 & 0 & 0 & f_{N1}f_{11}^*& f_{N1}f_{1N}^*& 0 & f_{NN}f_{11}^*& f_{NN}f_{1N}^*\\
0 & 0 & 0 & f_{11} & 0 & 0 & f_{1N} & 0 & 0 \\
0 & 0 & 0 & 0 & f_{11}f_{N1}^*& f_{11}f_{NN}^*& 0 & f_{1N}f_{N1}^*& f_{1N}f_{NN}^*\\
0 & 0 & 0 & 0 & |f_{11}|^2& f_{11}f_{1N}^*& 0& f_{1N}f_{11}^*& |f_{1N}|^2\\
\end{array}\right)\,,
\label{P}
\end{equation}
where the rows and columns are in the order
$00$, $01$, $0N$, $10$, $11$, $1N$, $N0$, $N1$, $NN$.

Inserting $\mathcal{P}$ into Eq.~(\ref{IP}), we obtain
\begin{eqnarray}
I_r(t)=\left|\sum_{j=2}^{N-1}\langle j|e^{-iHt}|1\rangle\langle
N|e^{iHt}|j\rangle\right|^2+\left|\sum_{j=2}^{N-1}\langle
j|e^{-iHt}|N\rangle\langle 1|e^{iHt}|j\rangle\right|^2\nonumber\\
+|f_{N1}f_{NN}^*|^2+|f_{NN}f_{N1}^*|^2+|f_{11}f_{1N}^*|^2+|f_{1N}f_{11}^*|^2
\end{eqnarray}
\end{widetext}

for the reduced interference.
This expression can be further simplified by using \begin{equation}
\sum_{j=1}^{N}\langle N|e^{iHt}|j\rangle\langle j|e^{-iHt}|1\rangle= \langle
N|1\rangle=0\,,
\end{equation}
such that
\begin{eqnarray}
I_r(t)=2\left|f_{11}f_{N1}^*+f_{N1}f_{11}^*\right|^2
+2|f_{11}f_{1N}^*|^2+2|f_{N1}f_{NN}^*|^2\,.
\end{eqnarray}
This result is valid for any Hamiltonian of the entire chain that
conserves the number of excitations. In the case of a linear chain
with symmetrical nearest-neighbor interactions (i.e. the Hamiltonian
is invariant under relabeling the qubits $1,2,..,N$ into $N,...,2,1$),
we have $f_{1N}=f_{N1}$ and $f_{11}=f_{NN}$. The reduced interference
can then be expressed using two amplitudes of the state transfer,
\begin{eqnarray}
I_r(t)&=&8|f_{11}|^2|f_{1N}|^2+2f_{11}^2(f_{1N}^*)^2
+2f_{N1}^2(f_{11}^*)^2\nonumber\\
&=&4|f_{11}|^2|f_{1N}|^2(1+2\cos^2(\gamma_{11}-\gamma_{1N}))\,,
\label{IPfin}
\end{eqnarray}
where $\gamma_{ij}=\arg(f_{ij})$. We are now in the position to
evaluate $I(t)$ for specific examples.

\subsection{Chains that conserve the number of excitations}
\label{sec.Bosechain}
Let us first consider the spin chains studied in \cite{Bose03}. They
consist of a one-dimensional array of $N$ spins, with nearest-neighbor
spins coupled through an isotropic Heisenberg interaction. The
Hamiltonian of the chain reads
\begin{equation} \label{Hbose}
H=-\sum_{i=2}^{N}J\bm{\sigma}_i\cdot\bm{\sigma}_{i-1}-
\sum_{i=1}^NB_i\sigma_i^z\,,
\end{equation}
where $\bm{\sigma}_i=(\sigma_i^x,\sigma_i^y,\sigma_i^z)$ denotes the
vector of the Pauli matrices on site $i$, $B_i$ denotes the
site-dependent static magnetic field and $J>0$ is
the coupling strength, taken as constant for all spins.

Bose showed that the average fidelity, Eq.~(\ref{ft}), for this model is
given by
\begin{equation} \label{Ft2}
F=\frac{|f_{1N}|\cos\gamma_{1N}}{3}+\frac{|f_{1N}|^2}{6}+\frac{1}{2}\,.
\end{equation}
At $t=0$, $f_{1N}=0$ for $N>1$, such that the average fidelity
corresponds to the fidelity of a random guess of Bob of the quantum
state of Alice ($F=1/2$).  The overlap $f_{1N}(t)$ becomes
appreciable, once the spin wave excited at Alice's end arrives at
Bob's spin. Perfect state transfer for all states ($F=1$) requires
$f_{1N}=1$, along with $\cos\gamma_{1N}=1$. The last equality can
always be achieved by varying the magnetic fields $B_i$.  From here on
we will assume that this is the case and therefore put
$f_{1N}=|f_{1N}|$ when plotting fidelities of the state transfer.

By comparing Eqs.~(\ref{IPfin}) and (\ref{Ft2}) one can see that interference
is determined by one more complex variable $f_{11}$ compared to the fidelity.
Therefore, in general there is no explicit formula that describes interference
in terms of fidelity alone. Naively one might expect that interference should
play an important role for quantum state transfer, if the fidelity of the
process exceeds the maximal classical value, $F=2/3$ \cite{Horodecki99}.
However, note that an ideal quantum state transfer can be realized through the
permutation of the first and the last spins
$|0\rangle_r\leftrightarrow|0\rangle_r$, $|1\rangle\leftrightarrow|N\rangle$,
which does not lead to any interference at all. In general, interference
measures both the equipartition of all output states for any computational
basis state as input, and the coherence of the propagation. ``Coherence'' was
defined in \cite{Braun06} as the sensitivity of the final probabilities
$\rho^\prime_{ii}$ to the initial phases. As is evident from Eq.~(\ref{rho'}),
the only phase information which contributes to the reduced interference in the
propagation through the spin chain is the relative phase between the states
$|0\rangle_r$ and $|N\rangle$. However, the coherence of the propagation
becomes irrelevant for perfect transfer, $f_{1N}=1$, as then $f_{NN}=0$ due to
the conservation of the number of excitations, and then the final probabilities
do not depend on any initial phases anymore. I.e.~for ideal state transfer, the
dynamics of the chain indeed realizes the above permutation with vanishing
interference. This is also evident from Eq.~(\ref{IPfin}) for
$f_{11}=f_{NN}=0$. Note, however, that the interference is finite during the
propagation of the signal through the chain, as well as quite generally for any
situation in which neither $f_{11}$ nor $f_{1N}$ vanish. All one can say is
that for $F(t)$ close to 1, i.e. $f_{1N}$ close to 1 and thus $f_{NN}$ close to
0, $I_r(t)$ remains quite small.

\begin{figure}
\begin{center}
\includegraphics[width=0.45\textwidth]{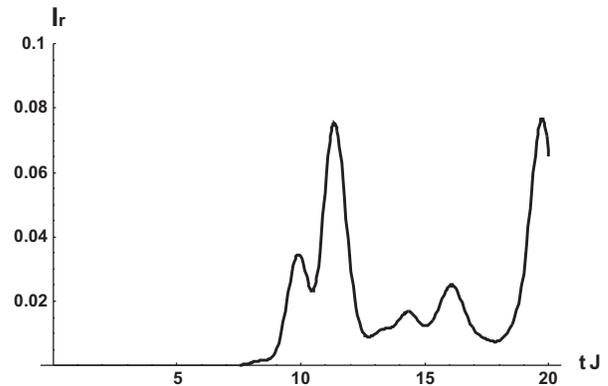}
\caption{Reduced interference for a spin chain with $N=20$ qubits
  described by the model defined in Eq.~(\ref{Hbose}).}
\label{fig1}
\end{center}
\end{figure}

Figure \ref{fig1} shows $I_r(t)$ that was obtained by numerically propagating
$|\Psi(t)\rangle$ for $N=20$ (see Eq.~(\ref{psit})) with the Hamiltonian
(\ref{Hbose}). The results are plotted with time in units of $1/J$. We also
assumed that $B_i=B$ for all $i$ and therefore the interference does not depend
on magnetic field. Indeed, in our model $B$ influences only the phases of
$f_{11}$ and $f_{1N}$ through a term $\exp(-2iBt)$ (see, for example
\cite{Bose03}) and according to Eq.~(\ref{IPfin}) the interference depends only
on phase differences and not on a global phase. One can see that the
interference remains quite small. This is because the probability to find an
excitation inside the chain is high and both quantities $|f_{11}|^2$ and
$|f_{1N}|^2$ cannot be big ($\sim 0.5$) at the same time at the time scale
that is relevant for quantum state transfer.

Let us now consider the case of reduced coupling constants of spins
$1$ and $N$ to the rest of the chain
\begin{equation} \label{Hreduced}
H=-\sum_{i=2}^{N}J_{i}(\sigma_{i}^{x}\sigma_{i-1}^{x}+
\sigma_{i}^{y}\sigma_{i-1}^{y})\,,
\end{equation}

\begin{figure}
\begin{center}
\includegraphics[width=0.45\textwidth]{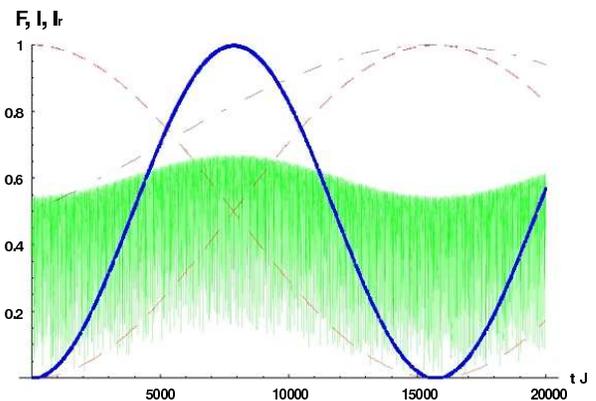}
\caption{Reduced interference $I_r$ for the case of small coupling
constants between
the first and the last pair of qubits, $N=8$ (blue, thick solid). The green
(thin solid) line shows the full interference of the entire chain (renormalized
by a factor $1/N$). The black (dot-dashed) line shows the fidelity $F(t)$. The
red (dashed) lines are the probabilities to find an excitation on the first
qubit and on the last qubit ($|f_{11}|^2$ and $|f_{1N}|^2$, respectively) if we
start with the excitation on the first qubit.} \label{fig2}
\end{center}
\end{figure}
with $J_{2}=J_{N}=aJ$ where $a\ll 1$, and $J_{i\neq 2,N}=J$.  It was
shown in \cite{Wojcik05} that this can drastically increase the
fidelity of the state transfer. Figure \ref{fig2} shows the reduced
interference $I_r(t)$ together with $F(t)$. We see that both are
perfectly anticorrelated. In particular, we have again $I_r(t)\simeq
0$ for $F(t)\simeq 1$ for the same reasons as discussed before. The
interference is maximal half way through the perfect state
transfer. In this case, the interference is not small (compare
Fig.~\ref{fig1} and Fig.~\ref{fig2}) since due to the weak coupling
the intermediate spins are only slightly excited \cite{Wojcik05} and
both quantities $f_{11}$ and $f_{1N}$ can be big ($1/\sqrt{2}$) at the
same time (see red (dashed) curves in the Fig.~\ref{fig2}).

\subsection{Chains that do not conserve the number of excitations}

\begin{figure}
\begin{center}
\includegraphics[width=0.5\textwidth]{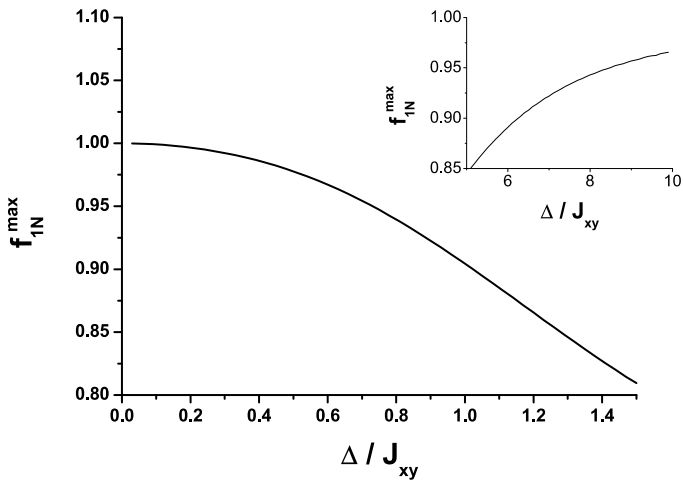}
\caption{Global maxima of $f_{1N}=\langle 1|e^{-iHt}|N\rangle$ in the time
interval [0,$1/J_{xy}$] as a function of $\Delta$
for the model defined by Eq.~(\ref{hamiltonian}), $N=3$.}
\label{fig4}
\end{center}
\end{figure}

\begin{figure}
\begin{center}
\includegraphics[width=0.5\textwidth]{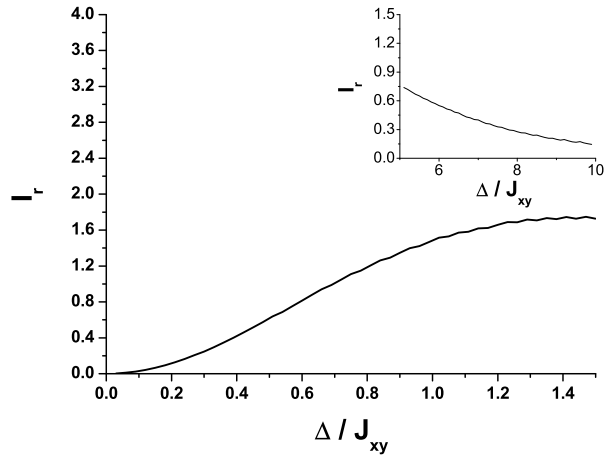}
\caption{Reduced interference at the global maxima of $f_{1N}=\langle
  1|e^{-iHt}|N\rangle$ in the time
interval [0,$1/J_{xy}$], see Fig.
\ref{fig4}.} \label{fig5}
\end{center}
\end{figure}

Now we consider a more general Hamiltonian that does not conserve the number of
excitations,
\begin{eqnarray}
H&=&-\sum_{i=2}^{N}[J_{xy}(\sigma_{i}^{x}\sigma_{i-1}^{x}+
\sigma_{i}^{y}\sigma_{i-1}^{y})+J_{z}\sigma_{i}^{z}\sigma_{i-1}^{z}]\nonumber\\
&-&
\sum_{i=1}^{N}(\Delta\sigma_{i}^{x}+B\sigma_{i}^{z})\,. \label{hamiltonian}
\end{eqnarray}
Equation (\ref{hamiltonian}) is a more realistic model than Eq.~(\ref{Hbose})
since in real qubits the $\sigma^x$ term, which describes the tunneling between
the states $|0\rangle$ and $|1\rangle$, cannot always be neglected. A physical
realization of Hamiltonian (\ref{hamiltonian}) was proposed in
\cite{Lyakhov05}. Sometimes the $\sigma^x$ term can be suppressed
\cite{Levitov,Lyakhov05}, but for longer chains even small values of $\Delta$
will influence the dynamics of the chain. In this case, Eq.~(\ref{IPfin}) is
not valid anymore, and the question of how much interference is used in the
quantum state transfer needs to be reassessed. Since the number of excitations
is not conserved, we have to do the calculation in the much larger Hilbert
space with dimensionality $2^N$ instead of $N+1$. One can numerically evaluate
$\rho'(t)$ and find $I_r(t)$ as a function of time. We used realistic qubit
parameters that are typical for flux qubits, see \cite{Orlando99, Lyakhov05},
namely $J_{xy}=0.08 J_z$, and $B=0$. The results of the calculations are shown
in Figs. \ref{fig4} and \ref{fig5}.

Figure \ref{fig4} shows the global maxima of $f_{1N}=\langle
1|e^{-iHt}|N\rangle$ in the time interval [0,$1/J_{xy}$] as a function
of $\Delta$ for a chain with $N=3$ qubits. The quantity $f_{1N}$
decreases with $\Delta$ until the time required for the state to be
transferred from the first to the last qubit is approximately equal to
$1/\Delta$. This is in agreement with \cite{Lyakhov05}. For large
$\Delta$, $f_{1N}$ is close to one due to excitations that are created
in the chain during this time interval.

Figure \ref{fig5} shows the reduced interference at the global maxima
of $f_{1N}=\langle 1|e^{-iHt}|N\rangle$ in the time interval
[0,$1/J_{xy}$]. Once again, interference decreases with increasing
$f_{1N}$, and vice versa, but this time as a function of the parameter
$\Delta$.  For very small $\Delta$, we have nearly perfect state
transfer (almost no equipartition and coherence), therefore the
interference is small. It increases with $\Delta$, as the creation of
excitations enhances the equipartition and sensitivity of the final
state to the initial state. For large $\Delta$, when a high value
$f_{1N}$ is achieved due to excitations created in the chain,
interference is small. For example if the excitation is created on the
last qubit, then the amplitude $|f_{1N}|$ will be equal to one. It
corresponds to nearly stochastic transfer, since the final
probabilities to find the last qubit in the state $|0\rangle$ or
$|1\rangle$ are almost independent of the initial state.

\section{Interference in the unitary propagation of the entire chain}
\label{section_full interference}
The result that perfect quantum state transfer is possible (and realized!)
without quantum interference is rather counter-intuitive. It is natural to
wonder what happens within the chain. Let us therefore open the black box and
study the interference in the propagation of the state of the entire chain
(called ``full interference'' $I=I_U/(D-1)=I_U/N$ in the following, where
confusion is possible) for chains which conserve the number of excitations.
This corresponds to a unitary propagation, and we will therefore employ
Eq.~(\ref{IU}) to quantify the interference.

\subsection{Chain with uniform coupling constants}
For a simple chain that consists of more than 3 qubits, the fidelity is
always less than one (except the case of specially engineered coupling
constants). This is due to the fact, that the the input state gets
dispersed over the spins at all times $t>0$.

Using the theory described in \cite{Lyakhov05} we calculated the eigenstates
and the eigenenergies of a more general version of the
Hamiltonian (\ref{Hbose}),
\begin{equation}
H=-\sum_{i=2}^{N}[J_{xy}(\sigma_{i}^{x}\sigma_{i-1}^{x}+
\sigma_{i}^{y}\sigma_{i-1}^{y})+J_{z}\sigma_{i}^{z}\sigma_{i-1}^{z}]-
\sum_{i=1}^{N}B\sigma_{i}^{z}\,. \label{HLyakhov}
\end{equation}

\begin{figure}
\begin{center}
\includegraphics[width=0.45\textwidth]{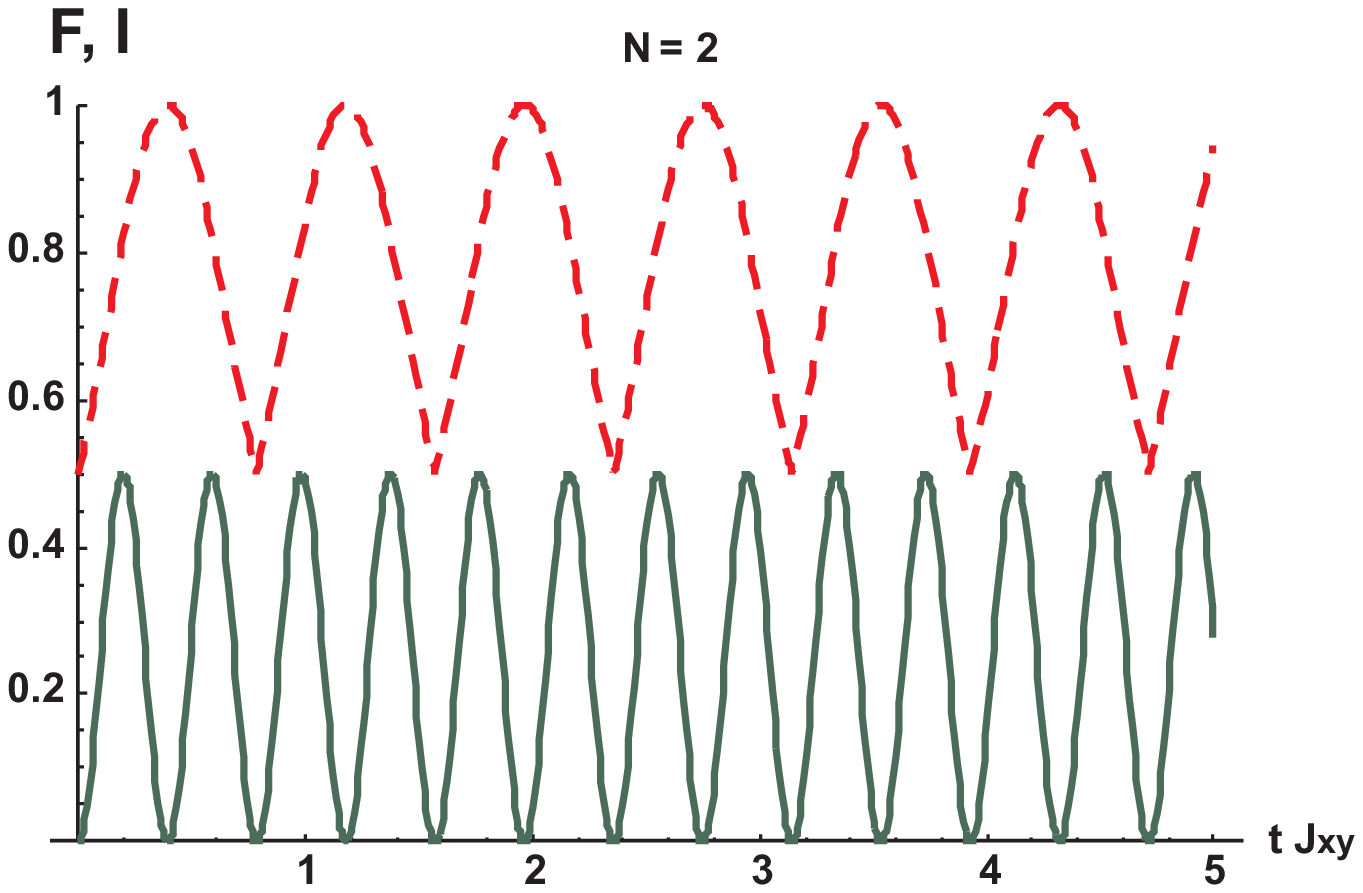}
\includegraphics[width=0.45\textwidth]{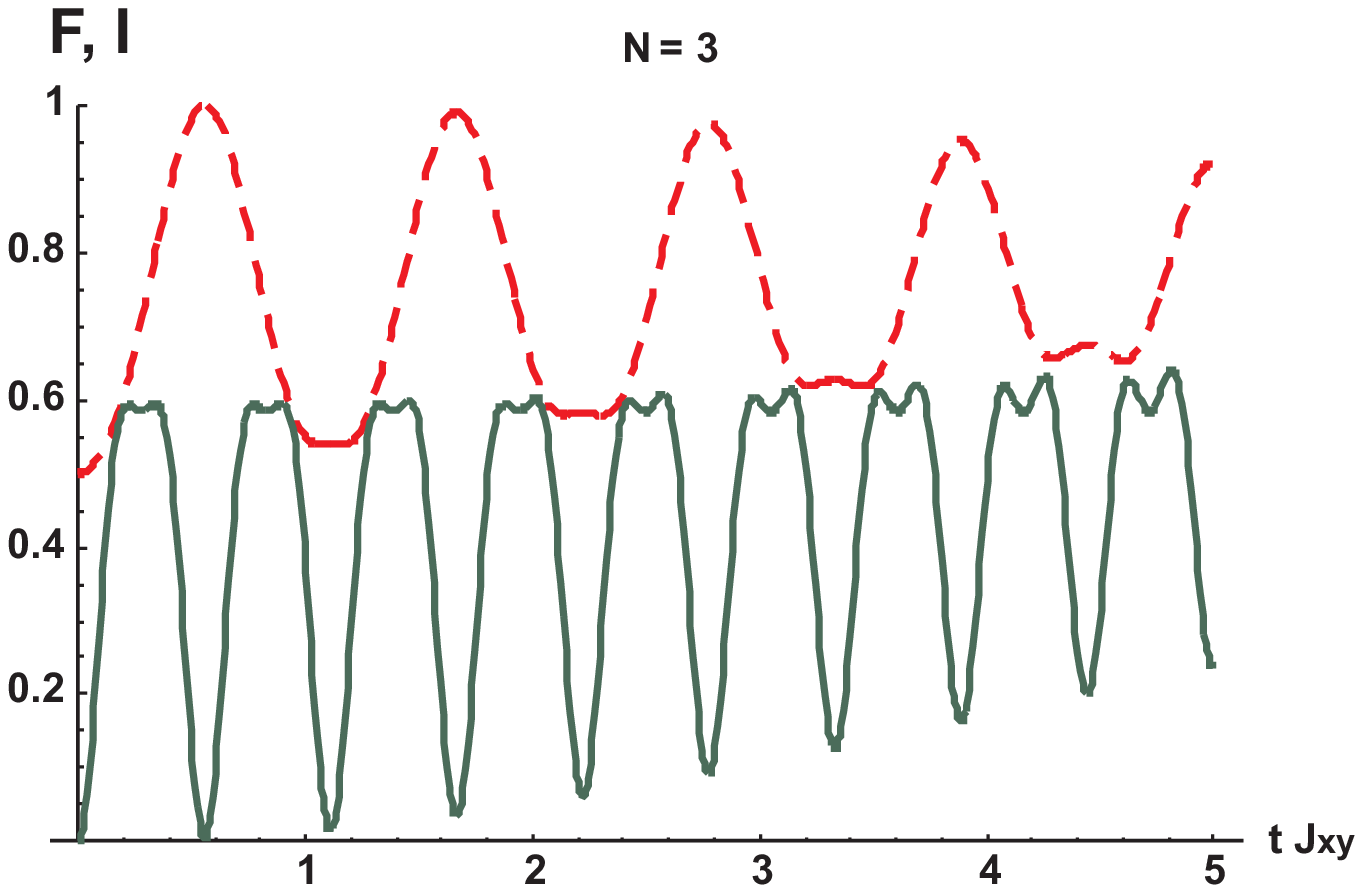}
\caption{Fidelity (red, dashed) and normalized full interference
  (green, solid) for
$N=2$ and $N=3$ qubits with uniform coupling constants, Eq.~(\ref{HLyakhov}).}
  \label{fig1tot}
\end{center}
\end{figure}

\begin{figure}
\begin{center}
\includegraphics[width=0.45\textwidth]{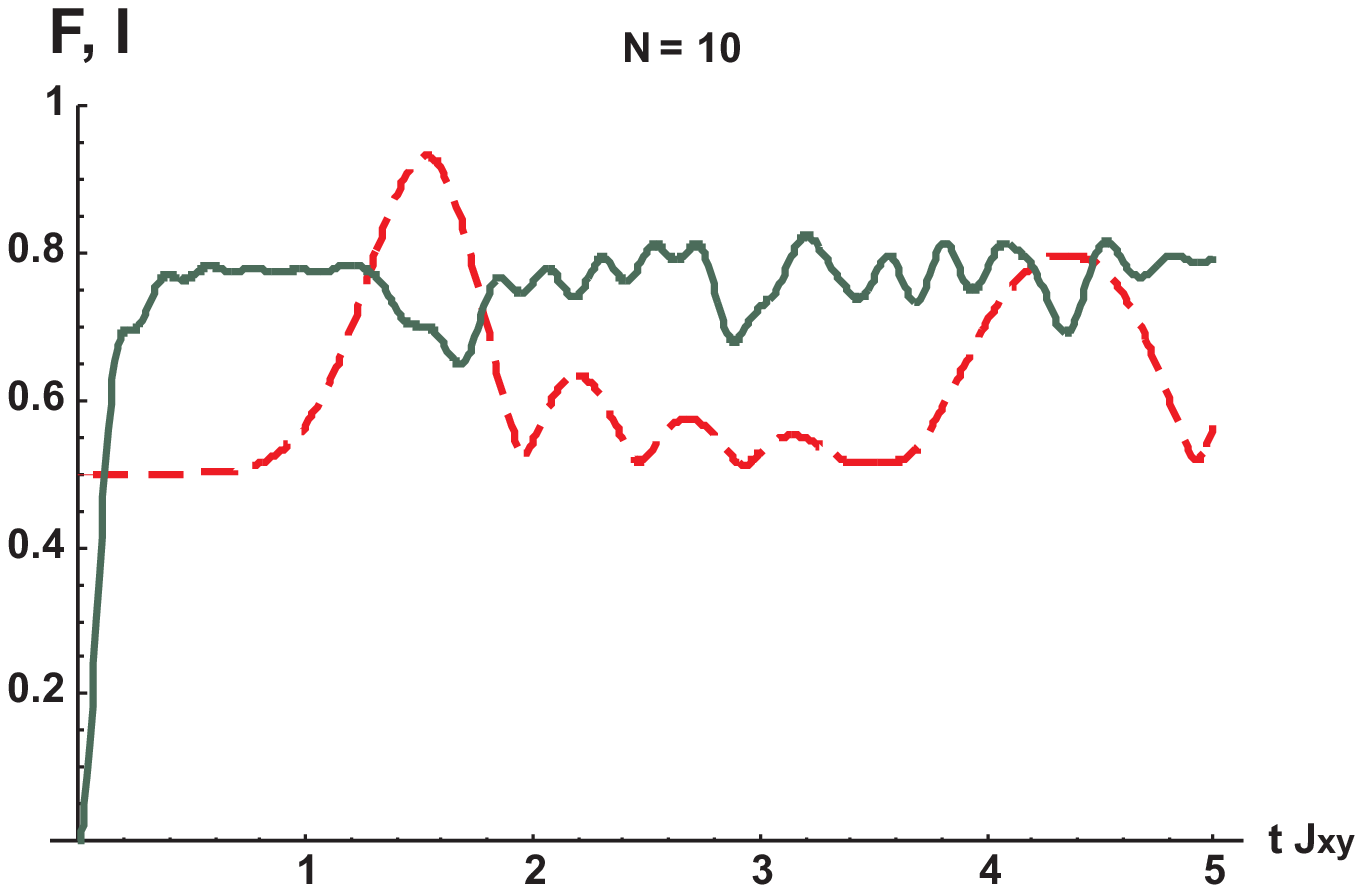}
\includegraphics[width=0.45\textwidth]{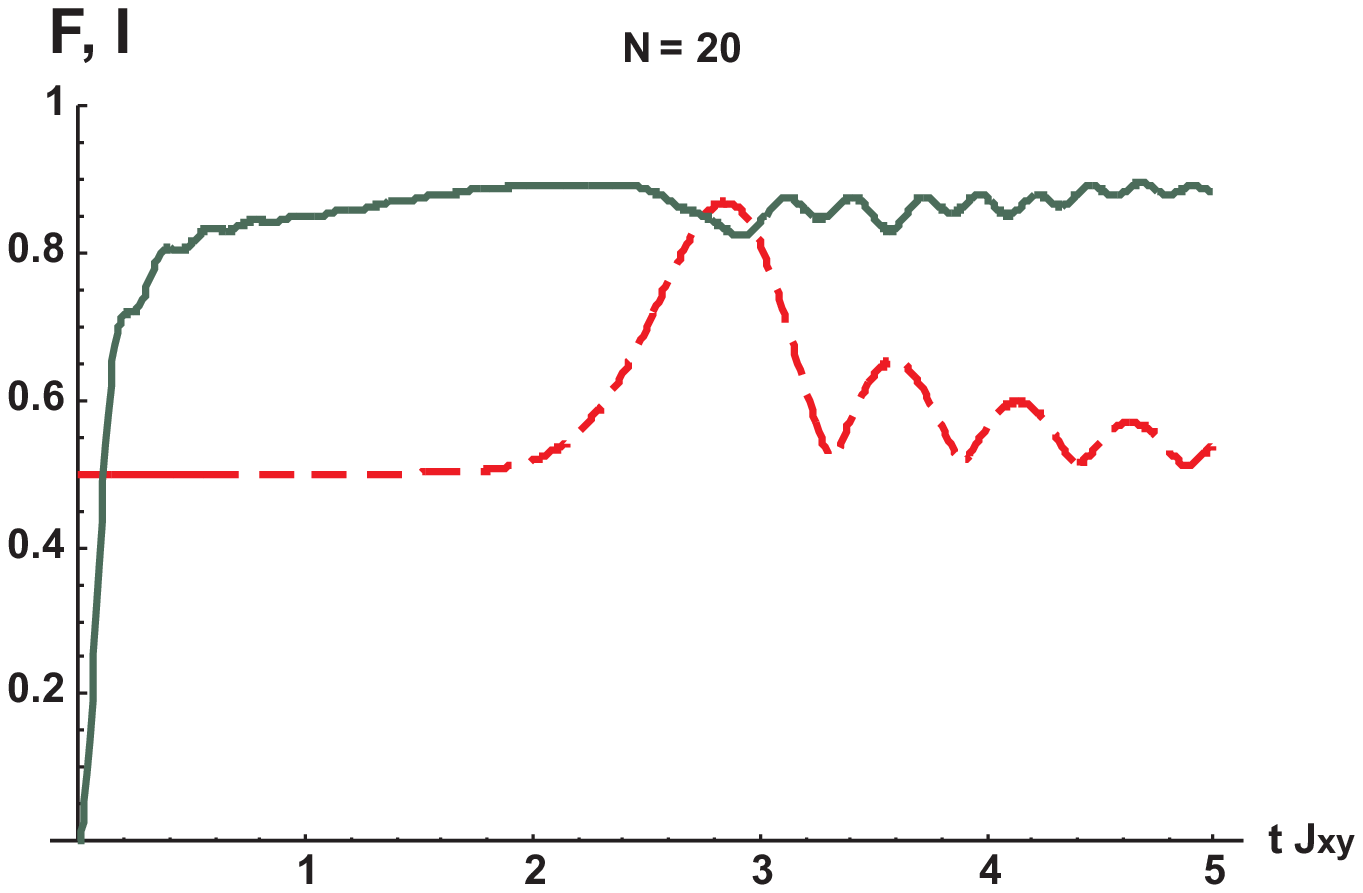}
\caption{Fidelity (red, dashed) and normalized full interference
(green, solid) for
$N=10$ and $N=20$ qubits with uniform coupling constants,
Eq.~(\ref{HLyakhov}).}
\label{fig2tot}
\end{center}
\end{figure}

This Hamiltonian also conserves the number of excitations and describes the
chains of superconducting qubits, proposed in \cite{Lyakhov05} and
\cite{Romito05}. Knowing the eigenvalues and eigenenergies of (\ref{HLyakhov})
allows us to find the matrix elements $U_{ik}$ and numerically calculate the
full interference as a function of time and of the number of the qubits in the
chain, restricting ourselves again to the $(N+1)$-dimensional Hilbert space of
the states in Eq.~(\ref{psin}). The results of these calculations are shown in
Figs.~\ref{fig1tot},~\ref{fig2tot} for
$J_z/J_{xy}=0.05$ \cite{Lyakhov05}.

\begin{figure}
\begin{center}
\includegraphics[width=0.45\textwidth]{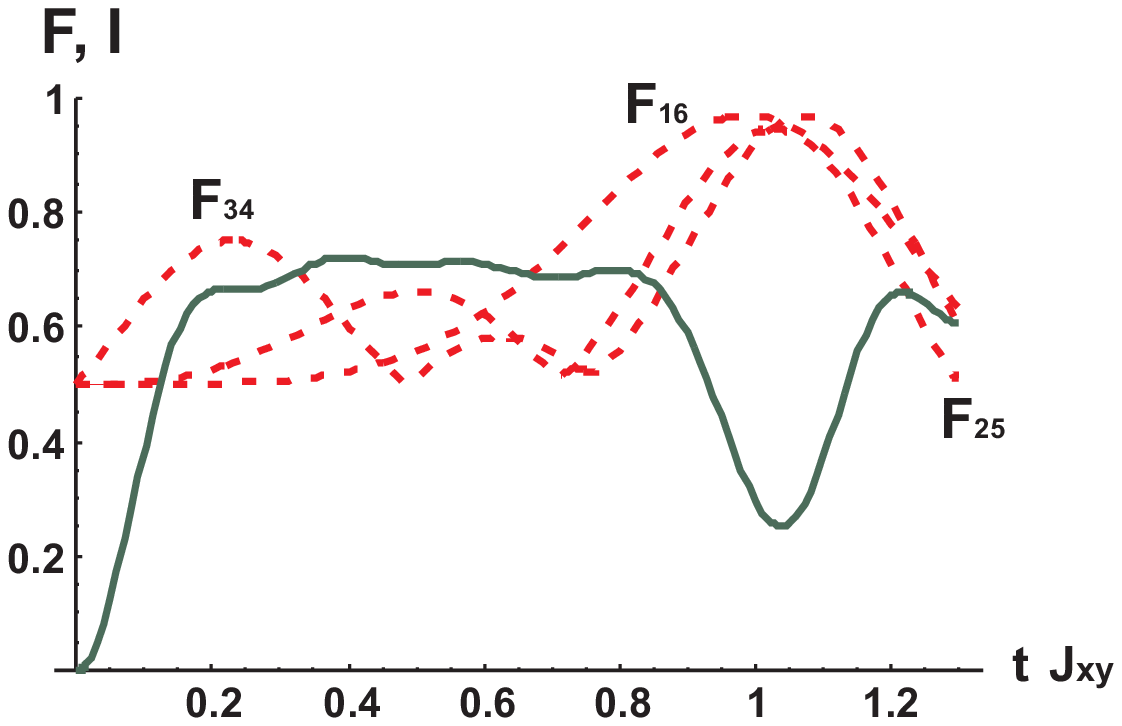}
\caption{Fidelities (red, dashed) and normalized full interference
(green, solid) for
$N=6$ qubits with uniform coupling constants, Eq.~(\ref{HLyakhov}).}
\label{figN6}
\end{center}
\end{figure}

\begin{figure}
\begin{center}
\includegraphics[width=0.45\textwidth]{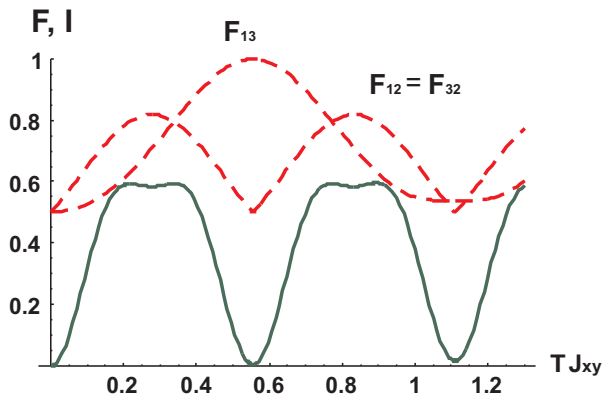}
\caption{Fidelities (red, dashed) and normalized full interference
(green, solid) for
$N=3$ qubits with uniform coupling constants, Eq.~(\ref{HLyakhov}).}
\label{figN3}
\end{center}
\end{figure}

As we can see in Fig.~\ref{fig1tot}, the full
interference is close to zero if the fidelity is close to one. The reason is
that the time that is required for
the excitation to be transferred from the first to the last qubit (the time of
the first fidelity maximum) is approximately equal to the time that it takes
for the excitation to travel from qubit $2$ to the end of the chain and then
back to qubit $N-1$ (and so on). This is illustrated in Fig.~\ref{figN6}, where
the fidelities $F_{ij}=\frac{|f_{ij}|}{3}+\frac{|f_{ij}|^2}{6}+\frac{1}{2}$ and
the normalized full interference are shown for the chain of $N=6$ qubits. We
can see that the interference is minimal in the region where local maxima of
the fidelities are located. When all maxima are close to one, then,
independently of the initial state, the final state will have
small equipartition and therefore the interference is small.

When the fidelity maximum goes down, the corresponding full
interference increases rapidly (see, for example, Fig.~\ref{fig1tot},
$N=3$). Hence, $I$ is very sensitive to the amplitude distribution of
the final state over the qubits. Here the amplitude of the spin $j$ in
the final state is $f_{1j}=\langle 1|e^{-iHt}|j\rangle$.

For short chains the fidelity maxima correspond to minimal
dispersion. For longer chains, the minima of the full interference are
shifted with respect to the fidelity maxima. This is due to that fact
that a maximal amplitude of the state ``up'' of the last qubit does
not necessarily correspond to the minimal dispersion as measured by
interference, which takes into account all possible input states.

Another feature of the interference graph are intermediate minima which
correspond to a partial localization of the excitation on the intermediate
qubits. For example in Fig.~\ref{fig1tot} ($N=3$) there are clear shallow local
interference minima that correspond to localization of the excitation on the
nearest neighbor of the initial qubit, see Fig.~\ref{figN3}. Deep minima
correspond to localization of the excitation after the state is transferred
through the whole chain. For longer chains, the times when the excitation is
localized on intermediate qubits depends on the initial state, i.e.~the
fidelity maxima do not exactly coincide (see Fig.~\ref{figN6}). Therefore these
small features are less pronounced.

\subsection{Reduced coupling constants at the end of the chain}

The full (normalized) interference $I(t)$ for a chain with reduced coupling
constants between the first and the last pair of qubits is shown as
green solid line
in Fig.~\ref{fig2}. $I(t)$ oscillates rapidly on the time scale of the
state transfer from the first to the last qubit, with an envelope whose
upper boundary
perfectly correlates with the oscillations of the reduced interference
and an amplitude which is, for $N=8$, about a tenth of the amplitude of the
reduced interference $I_r(t)$. This behavior is indeed to be expected from the
fact that $I(t)$ is a sum of equipartition measures for all initial states
localized on any qubit in the chain, whereas $I_r(t)$ measures equipartition
only on the first and last qubit. As the state transfer is basically perfect
(and therefore $I_r(t)=0$ for $t=0$ and at the time of optimal transfer
$t=t_1$), the lack of this contribution to the equipartition measure leads to a
minimum in the envelope of $I(t)$ at $t=0$ and $t=t_1$. At the same time, the
maximum of the envelope of $I(t)$ halfway through the state transfer
(corresponding to an additional contribution of about 0.1 to $I(t)$) indicates
that the equipartition of the first and last qubit captures the essence of the
equipartition in the chain for an initial state localized on the first qubit.
This agrees with Ref.~\cite{Wojcik05} since there is only a small amplitude for
an excitation inside the chain during the state transfer. Therefore the
equipartition between the first and the last qubit gives the main contribution
to the full interference.

\section{Conclusions}
In summary we have calculated the interference during the transfer of a quantum
state through several types of one-dimensional spin chains with
time-independent nearest-neighbor coupling constants, both for chains which do
or do not conserve the number of spin excitations. We have shown that for a
high-fidelity transfer the reduced interference of the propagator of just the
first and last qubits is very small, and vanishes for perfect transfer. This
can be understood from energy conservation and the fact that interference
measures, besides phase coherence, the equipartition of the final states for
all computational states taken as input states. The full interference of the
entire chain (propagated unitarily) shows rapid oscillations on the time scale
of a complete transfer. For a chain with reduced coupling constants between the
first and the last pair of qubits the envelope of these oscillations follows
the reduced interference. Thus, interference is not only valuable tool for
investigating quantum algorithms, but also gives us a deeper insight into the
dynamics of quantum state transfer.

{\em Acknowledgments:} This work was supported by the Agence National de la
Recherche (ANR), project INFOSYSQQ, the EC IST-FET project EuroSQUIP, the Swiss
NSF, and the NCCR Nanoscience.

%\end{twocolumn}
\end{document}